# Privacy of Big Data in the Internet of Things Era

*Charith Perera (Australian National University), Rajiv Ranjan (CSIRO Digital Productivity Flagship), Lizhe Wang (Chinese Academy of Sciences), Samee U. Khan (North Dakota State University), and Albert Y. Zomaya (University of Sydney)*

## Abstract

Over the last few years, we have seen a plethora of Internet of Things (IoT) solutions, products and services, making their way into the industry's market-place. All such solution will capture a large amount of data pertaining to the environment, as well as their users. The objective of the IoT is to learn more and to serve better the system users. Some of these solutions may store the data locally on the devices ('things'), and others may store in the Cloud. The real value of collecting data comes through data processing and aggregation in large-scale where new knowledge can be extracted. However, such procedures can also lead to user privacy issues. This article discusses some of the main challenges of privacy in IoT, and opportunities for research and innovation. We also introduce some of the ongoing research efforts that address IoT privacy issues.

## Introduction

The Internet of Things (IoT) [1] is a network of networks, in which, typically, a massive number of objects/things/sensors/devices are connected through the information and communications infrastructure to provide value-added services. The IoT allows people and things to be connected Anytime, Anyplace, with Anything and Anyone, ideally using Any path/network and Any service [2]. Additional definitions on IoT are also listed in [2]. It is predicted that, by 2020, there will be 50 to 100 billion devices connected to the Internet [2]. These devices will generate Big Data [3] that needs to be analysed for knowledge extraction. Even though data collected by individual devices may not provide sufficient information, aggregated data from number of physical devices and virtual sensors (e.g. social media such as Facebook, Twitter) [2] can provide a wealth of knowledge for important application areas including disaster management, customer sentiment analysis, smart cities, and bio-surveillance.

There is no clear definition for Big Data [3]. It is defined based on some of its characteristics. The big data does not mean the size. There are three characteristics that can be used to define big data, as also known as 3V's: volume, variety, and velocity [4]. Volume relates to size of the data such as terabytes (TB), petabytes (PB), zettabytes (ZB), etc. Variety means the types of data. In addition, difference sources will produce big data such as sensors, devices, social networks, the web, mobile phones, etc. Velocity means how frequently the data is generated (e.g. every millisecond, second, minute, hour, day, week, month, year).

The collection and analysis of data in the IoT applications has many objectives. For example in case of customer sentiment analysis, such data can be used for improving personalized recommendations hence leading to better customer experiences. On other hand in case of smart cities, governments and city councils can use the knowledge extracted to make strategic decisions (e.g., placement of traffic lights, construction of new roads/bridges, etc.) and future city plans [5], [6]. However, the data collected by smart IoT devices may contain very sensitive personal data based on type of application and data sources. Therefore, such data must be managed carefully to avoid any user privacy violations. Consequently, in the subsequent text, we briefly discuss the importance on addressing IoT privacy challenges [7]. Users are the people or the consumers who are using the product or the service.

In rest of this paper, we briefly introduce privacy in IoT, followed by different survey results that consolidate public opinion towards IoT solutions and their impact towards user privacy. Later, we discuss a number of research challenges that need to be addressed over the coming years to make sure the IoT will become a pleasant experience to the users as well as to the business. Finally, we highlight some leading research and development efforts in IoT and privacy domains.

## Privacy in Action: Looking Back

Golbeck and Mauriello [9] have shown that the average Facebook users significantly underestimate the amount of data they that they give access to third party applications. Moreover, they also noted that most of us tend to overlook the privacy [8] terms and policies on the Web.

In the IoT era, the amount of user data that can be collected can be significantly higher. For example, recent wearable technologies, such as Google Glass, Apple iWatch, Google Fit, Apple Health Kit, and Apple Home Kit may collect very sensitive information about users, ranging from their health conditions to financial status by observing/recording daily activities. It is noteworthy to mention that to succeed in the IoT marketplace, product and service providers need to gain the consumer confidence [10].

We note that privacy issues during the Internet age did receive significant attention over the last few years. For example, *'allegation of governments spying on their citizens'* to the new laws such as *'right to be forgotten'* [11] has opened up a whole range of debate. Compared to the Web era, the IoT is more vulnerable to privacy violations. Therefore, researchers as well as IT professionals will pay more attention towards IoT technologies, business models, and potential regulatory efforts to ensure that a more secure and privacy preserved IoT data management techniques are developed.

TRUSTe [12] highlighted the fact that privacy concerns could be a significant barrier to the growth of IoT. According to the TRSUTe survey, about 60% of internet users have basic privacy awareness of IoT and they know that smart devices, such as smart TVs, fitness devices, and in-car navigation systems could collect personal activities data. Moreover, 85% of the Internet users would like to understand more about data collection. Furthermore, 88% of the respondents wanted to control the data collection from the smart devices. Finally, the survey revealed that 87% of internet users were concerned about the type of personal information collected.

## Trends, Predictions, and Opinions

Perera and Zaslavsky [13] conducted a survey to find out the public opinion about the Sensing as a Service model [5], which is an IoT business model for data markets. Sensing as a Service model envisions to create a market place for contextually enriched sensor data to be exchanged between different parities for financial or social benefits. The survey was conducted among 137 participants in the United States (US). The respondents were asked to assume 100% guaranteed privacy and security. According to the results, majority of the responses (64%) were in favour of trading-based Sensing as a Service model. In a market environment where owners of the IoT solutions can sell data, 67% of the respondents expected less than $500 worth of value returned per year. The survey also revealed that 66% of the respondents were happy to make a large initial investments and additional investments required to support the Sensing as a Service model, as long as the additional cost can be recovered within three months. Furthermore, the survey also revealed that Sensing as a Service model (also called Data Markets) motivated 65% of the respondents to purchase smart devices for IoT adaptation, even at higher prices than current market value. In a secondary survey, Perera and Zaslavsky [13] asked 1,000 US participants, whether they would like to exchange data for

a financial return without explicitly mentioning and assuring user privacy. As expected, 79% responded negatively to such an idea.

Fortinet [14] conducted a survey on the consumer interest towards the IoT marketplace focusing on the adaptation of the IoT devices by 1,801 homeowners. The survey was administered in Australia, China, France, Germany, India, Italy, Malaysia, South Africa, Thailand, United Kingdom, and the US. According to the survey, 61% of the homeowners agreed that the connected home is 'extremely likely' to become a reality in the next five years. At the same time, 68% of the respondents were 'extremely' or 'somewhat' concerned about the exposure of personal data. Around 57% of the respondents have considered privacy as an important issue in the IoT, and they currently do not understand or trust how the data collected though their IoT device would be used. According to Fortinet, 67% of the respondents consider data privacy as an extremely sensitive issue. Moreover, 70% of the homeowners said that they want to have personal control over the collected data, and 66% of the respondents believe that only they or those to whom they give permission should have access to the data. Almost half of the respondents (54%) also expected government or non-government organizations to regulate data collection and processing in the IoT domain to mitigate privacy issues. As regards to the question regarding the responsibility on the vulnerabilities of the IoT solutions, the respondents seemed to have a divided opinion, where 48% believed that the device manufacturer was responsible for updating/patching their devices, and 31% believed that it was the homeowner's responsibility. Fortinet's survey also found that the homeowners were willing to pay for a connected home, where 40% responded with 'definitely' and another 48% have said 'maybe'. It is evident that an innovative business model is required to motivate the 48% of the 'may be' group towards investing in the IoT solutions. This is where business models, such as Sensing as a Service would come into play.

## Privacy Challenges in the IoT

Today, consumers of online services are aware that when they use free online services (e.g., email, social networking, and news feeds), they automatically become the data sources of the business, who can analyse this data to improve customer satisfaction. Even worse, the data can be sold to third party for further analysis. However, in the future IoT era it is likely that service providers will adopt one of the two following models. First, some consumers may willingly pay for consuming the services with the aim of protecting their privacy. Second, some consumers may offer to give away data, of course under some limitations and conditions, in return for consuming services free of any charge.

Data collected through smart wearable and smart home devices can be used to generate contextually [2] enriched information. Device owners should remain in charge of such data at all-time despite they may give access to their data to external parties temporarily in order to accomplish a specific task. Consequently, the IoT era poses significant privacy challenges, especially due to sheer scale of the IoT. The EU Commission report on the IoT, CERP-IoT [10], has identified security and privacy as a major IoT research challenge, including: privacy preserving technology for heterogeneous device sets, models for decentralised authentication, trust, energy-efficient encryption, data protection technologies, security and trust for cloud computing, data ownership, legal and liability issues, repository data management, access and use rights, rules to share added value, responsibilities, liabilities, artificial immune systems solutions for IoT, secure, low cost devices, integration into, or connection to, privacy preserving, frameworks, and privacy policies management. In the subsequent text, we introduce and discuss some of the major IoT privacy challenges [15]. A summary of research questions are presented in the Supplemental Material section.

***User Consent Acquisition:*** In the IoT, user consent is about acquiring the required level of permission from the users and non-users who are affected by the devices or services. In the traditional Web, the method of receiving user consent is through the privacy terms and policies presented to the users through paragraphs of long text. Recently, with the emergence of social media and mobile apps, consent acquiring mechanisms have changed. Researchers [9] have found that the current methods of asking user consent in social media platforms, such as Facebook are ineffective and most of the users underestimate the authorization given to the third party applications. In some cases, developers may not provide accurate information to the users for the consenting decision. In other cases, developers may provide accurate information; however, the users would be unable to understand exactly what the consent entails for the lack of technical knowledge. One of the major privacy challenges in the IoT is to develop technologies that request consent from users in an efficient and effective manner. This is a challenging task due to the fact that every user has very limited time and limited technical knowledge to engage in the process. Such research will need to combine principles and techniques of the human computer interaction and cognitive sciences.

***Control, Customization, and Freedom of Choice:*** In the IoT, the data owner must have full control on data, allowing the users to delete or move data from one service provider to another at any time. Unfortunately, existing IoT solutions in the marketplace only provide a limited access to the users. Moreover, the users should be able to choose the hardware devices and software components from different vendors to build their smart environments (e.g. smart home). This gives full control and freedom of choice to the users. Consequently, the users must decide on what kind of data to be shared with what access rights to the service providers. Users should also be facilitated with the functionalities to withdraw or change pervious user consents. It is also important to understand that without having access to some types of data, a service provider will not be able to facilitate certain types of services. However, service providers must not unfairly treat consumers, such as by disabling curtain features to motivate users for providing consent, changes to the subscription fees, etc.

***Promise and Reality:*** Each of the IoT solution promises to offer a selected number of functionalities. This is achieved by the service providers by requiring certain types of raw data to be processed and analysed. However, with the development of new technologies, business may be able to derive more knowledge from the user acquired data. However, if the service providers want to use the raw data to derive more knowledge, then the providers must explicitly request permission by explaining the new possibilities and potential consequences to the users. The bottom line is that the service providers must not use the already collected data for any other purpose without explicit user consent. Both regulations and technology need to be developed and put in place to avoid such a misuse.

***Anonymity Technology:*** Network communication interfaces typically have MAC addresses that can be used to trace back the data communication paths. The combination of multiple MAC addresses of multiple devices will help create unique fingerprints and a unique profile where analytics can be used to extract knowledge. Consequently, user location can easily be tracked. It is important to discover new technologies that can anonymize data communication paths to protect user privacy. Due to the usage of large number of sensors and service, it is challenging to anonymise multi-dimensional data. Specially, it is easy to build fairly unique profiles that may enable knowledge extraction for a particular user. Currently, the network communication technologies do not preserve the anonymity of the users. Newer IoT platforms will be required to adopt technologies, such as Tor (torproject.org) that is a technology that conceals user location. In essence, a comprehensive anonymization framework is required to facilitate end-to-end anonymity in the IoT. Such a framework must ensure anonymity at different levels, such as data modelling, storage, routing, communication, analytics, and aggregation.

**Security:** Even though a detailed discussion on IoT security [16] is out of the scope this article, it is important to mention that standardization and certification would be the foundation of security. Moreover, all of the stakeholders have the responsibility to secure the infrastructure, the data collection and transfer process, as well as the people using the devices. Device manufactures will need to upgrade or patch the firmware and software. Moreover, such updates must be pushed onto the devices and must be automatically installed with minimum user intervention. It may also be the owner responsibility to make sure that their IoT devices and software systems remain up-to-date. Security need to be ensured throughout the data flow within the IoT.

The consequences of releasing or selling private data of users could result in users' receiving annoying customized target advertising. In more extreme circumstances, criminals may use such data to perform different types of criminal activities that could harm individual consumers (e.g. identifying user behavioural patterns to invade houses) or entire communities (e.g. identifying critical timeframes to destruct water supply or energy distribution channels).

## Stakeholder Responsibility

As depicted in Figure 2 (in Supplemental Material section), we identify five major stakeholders (described below), namely: *device manufacturers, IoT cloud services and platform providers, third party application developers, Government and Regulatory bodies, and Individual Consumers and non-consumers* [15].

**Device Manufacturers:** Device manufactures must embed privacy preserving techniques into their devices. Specially, manufactures must implement secure storage, data deletion, and control access mechanisms at the firmware level. Manufactures must also inform consumers about the type of data that are collected by the devices. Moreover, they must also explain what kind of data processing will be employed and how and when data would be extracted out of the devices. Next, the manufactures must also provide the necessary control for the consumers to disable any hardware components. For example, in an IoT security solution, consumers may prefer to disable the outside CCTV cameras when inside the home. However, consumers will prefer to keep both inside and outside cameras active when they leave the premises. Moreover, devices manufactures may also need to provide programming interface for third party developers to acquire data from the devices.

**IoT Cloud Services and Platform Providers:** It is likely that most of the IoT solutions will have a cloud based service that is responsible for proving advance data analysis support for the local software platforms. It is very critical that such cloud providers use common standards, so that the consumers have a choice to decide which provider to use. Users must be able to seamlessly delete and move data from one provider to another over time. Such a possibility can only be achieved by following a common set of interfaces and data formats. Most of the cloud services will also use local software and hardware gateways such as mobile phones that act as intermediator controllers. Such devices can be used to encrypt data locally to improved security and to process and filter data locally to reduce the amount of data send to the cloud. Such methods will reduce the possibility of user privacy violation that can occur during the data transmission.

**Third Part Application Developers:** Application developers have the responsibility to certify their apps to ensure that they do not contain any malware. Moreover, it is the developers' responsibility to ensure that they present clear and accurate information to the users to acquire explicit user consent. Some critical information are: (*1*) *the task that the app performs, (2) the required data to accomplish the tasks, (3) hardware and software sensors employed, (4) kind of aggregation and data analysis techniques that the app will employ, (5) kind of knowledge that the app will derive by data processing*.

Users need to be presented with a list of features that the application provides, and the authorization that the user needs to give to activate each of those features. The control must be given to the user to decide which feature they want to activate. Moreover, in the IoT, acquiring user consent should be a continuous and ongoing process. Consequently, the application developers must continuously allow the users to withdraw, grant, or change their consent. Moreover, users must be given full access to the data collected by the IoT devices.

***Government and Regulatory Bodies:*** Either government or independent regulatory bodies must lead and enforce standardization and legal efforts [17]. Standardization efforts should comprise both a certification process and a technology development process. However, such efforts must not hinder innovation but ensure interoperability among different IoT solutions, and fair marketplace and competition. Standardization of data transfer and storage will reduce the entry barriers to the IoT market place. For example, there are some standardization efforts going on within the IoT domain, such as AllSeen Alliance (allseenalliance.org) that has attracted a number of leading industries. It is important to establish a governing body similar to the W3C for the IoT to oversee the standardization and certification processes. Some of the critical areas for standardization would be: communication, device descriptions and discovery, data exchange, encryption, user consenting mechanisms, data, modelling, storage, and routing. As stated earlier, the standardization efforts must be complemented with a certification process. Currently, individual companies are attempting to certify devices and apps by themselves. Unfortunately, such efforts will hinder the interoperability. The certification mechanism for the IoT would be similar to the 'certificate authority model' that is used for the Internet. However, the IoT certification model would be much broader, as it may need to certify both hardware products and software services.

***Individual Consumers and Non-Consumers:*** The individual stakeholders can be both IoT product consumers and non-consumers. Most of the exiting IoT solutions are mainly focussed on consumers. However, non-consumers can also be affected by some kinds of IoT solutions. For example, IoT products such as Google Glass pose threat to not only to the wearers but also to the people within the viewpoint. The IoT device owner should be sensitive to matters similar to the above. Moreover, when the IoT devices are installed in private homes, office environments, or apartment complexes, it is important to notify the non-consumers regarding the nature of the solution deployed and related information. Such notification would be similar to a CCTV surveillance notification we see in public places today. However, due to the complexity of the monitoring and actuation tasks, it may necessary to employ interactive and digital means to inform the non-consumers.

## State of the Art: Academic Research to Start-ups

The OpenIoT (openiot.eu), an European Union funded project, is a IoT cloud platform that supports Sensing as a Service model. OpenIoT provides instantiations of cloud-based and utility-based sensing services enabling the concept of Sensing as a Service, via an adaptive middleware framework for deploying and providing cloud services. However, the OpenIoT does not adequately address privacy issues. Instead, OpenIoT promotes the use of public data sources, such as Linked Open Data.

Lab of Things (LoT) [18] by Microsoft Research is a flexible platform that uses connected devices in homes. It enables researchers to easily interconnect devices and implement application scenarios, and sharing of data, code and participants, which further lowers the barrier to evaluate ideas in a diverse set of homes. However, the LoT assumes that the privacy concerns must be manually handled where the deployer must sign an agreement with data owners.

Hub of All Things (HAT) (hubofallthings.com), Funded by the EPSRC, is an ongoing project that aims at developing data markets to support trading data generated by the IoT solutions in smart home environments. The HAT does not address privacy issues. Its primary goal is to provide an API so that the home owners can push data to the cloud.

There also are a number of industrial efforts to build IoT platforms. Xively (xively.com) offers a Platform as a Service that allows IoT devices to connect to the cloud. It does not address any privacy issues other than that it will provide secure data storage. In Xively, privacy protection is the responsibility of the person who builds applications and services using the Xively platform.

Datacoup (datacoup.com) is a new start-up that will allow users to sell personal data. Primarily, their focus is on social media data, such as Facebook, Twitter, and YouTube. Currently, Datacoup does not focus on IoT data. Datacoup is among the few initiative that focuses on trading any kind of personal data. Datacoup pays $8 for each user[1] that shares data. However, users have to trust Datacoup completely as Datacoup will sell user data through their own servers. Mydex (mydex.org) is a British social-enterprise helping to make it easier and safer for individuals to hold, control, and re-use their personal information in effective and secure ways. Mydex is also a personal data sharing platform.

## Conclusions

Collecting data through IoT solutions and analyse them in large-scale have a significant value to offer for both individual users and businesses. Further, it can also make significant impact towards society in general through increase productivity and reducing wastage. However, existing technologies and regulations are not sufficient to support privacy guaranteed data management life cycle. From the time the data is being captured by the sensors embedded in IoT solutions to the point where knowledge is extracted and raw data is be permanently and securely deleted, user privacy need to be protected and enforced. By doing this only the IoT solutions can gain the confidence of the consumers. Limitation of the technology will need to be mitigated by strict laws and regulation that would include strict and serious penalties for offenders and misusers.

Future research efforts will focus on developing novel efficient and scalable privacy preserving algorithms that efficiently scales across IoT data processing technologies (SQL/NoSQL datastores, batch processing systems, and stream processing systems) while adapting to uncertain data sizes and data variety. This will be achieved by exploiting the inherent workload and resource performance features of big data processing technology for scaling privacy preserving algorithms.

---

[1] http://www.technologyreview.com/news/524621/sell-your-personal-data-for-8-a-month/

**Supplemental Material**

| | |
|---|---|
| **User Consent Acquisition** | • How to present privacy policies and terms to the users in IoT in a user friendly and understandable manner? |
| *Control, Customization, and Freedom of Choice* | • How to allow users to control and mange their data?<br>• How to ensure interoperability between vendors to assure freedom of choice? |
| *Promise and Reality* | • How to ensure that service proders will not use data for any other purpose than what users have given permission to? |
| *Anonymity Technology* | • How to main tain the anonimity of the users through out the differnt phases of data life cycle? |
| Security | • How to protect the data (through out its life cycle) as well as the infrastructure from external forces with malicious intents? |

*Figure 1: Summary of Research Questions*

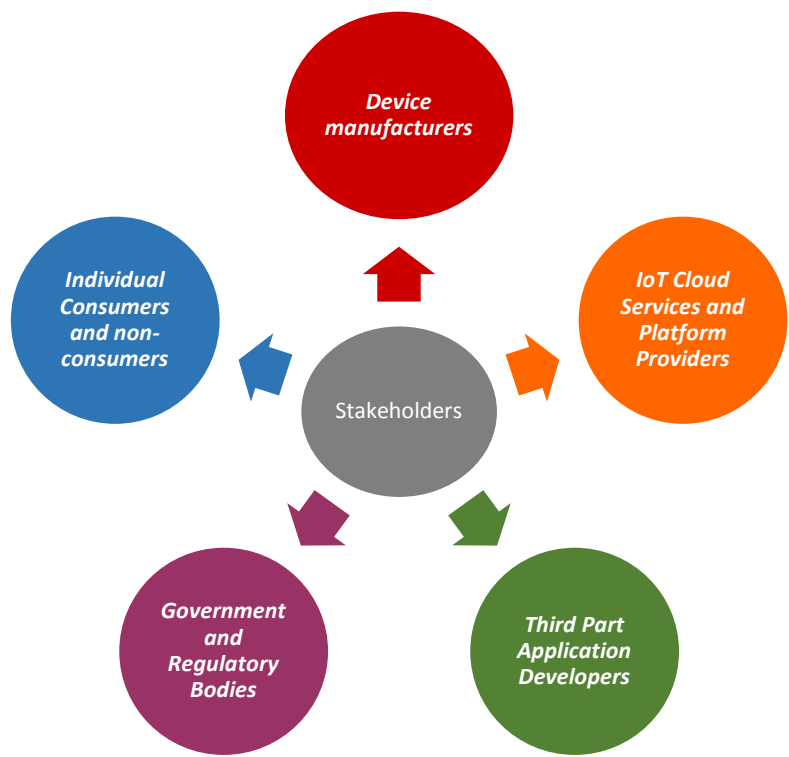

*Figure 2: Major Stakeholders Responsible for Protecting User Privacy*

| Traditional Computing Model (Grids, Supercomputers) | Cloud Computing |
|---|---|
| Fixed capacity | Elastic capacity |
| Mutual virtual organization based research community wide resource sharing | Offered under pay-as-you-go utility model |
| Limited adoption of virtualization technologies | Software and hardware virtualization technologies are key to the design and development of cloud computing datacentres |
| Driven by scientific applications such as high energy physics, cancer research, protein folding, computational drug design, kidney modelling, and astronomy | Driven by both scientific application from grid computing era as well as current generation commercial applications such as distributed gaming, social networking, web hosting, medical imaging, click stream analysis, credit card fraud detection, demand-response modelling for smart energy grids, disaster management |
| Lack of strict quality of service (QoS) guarantees in terms of availability and other metrics | Higher guarantees (99.99%) for availability, however guarantees for other QoS metrics based on application scheduling intelligence |
| Setup and managed by universities and large research consortium | Setup and managed by third party providers such as Amazon, Microsoft, and Rackspace |
| High upfront costs | High upfront costs for cloud datacentre provider but low upfront costs for application providers |

*Table I: Comparing traditional computing model and cloud computing.*

### State of the Art: Academic Research to Start-ups

Privacy preserving data analytical services, such as Aircloak (aircloak.com) collects data from multiple sources and brings them to a single location to perform analytics. We believe that an ideal platform must be able to analyse data in different locations (e.g. within the smart home or outside) and at a later stage perform aggregation [1]. Centralized data analysis creates a significant security risk, as well as the privacy risk due to potential misuse of raw data. It is evident that such platforms give only limited attention to the IoT privacy issues. One of the primary reasons is the immaturity of the IoT. Until recently, the main focus was on building IoT infrastructures.

Dataware [2] aims to develop infrastructure to store and process personal data within the household environment. User data is treated as immovable and the third-party applications are granted capabilities to run against the data wherever it is stored. Recently, the European Union funded project, User Centric Networking (UCN) (usercentricnetworking.eu), is an initiation to build privacy guaranteed content recommendation system based on personal data. The UCN is application oriented where it focus on facilitating an analytical function (i.e. recommendation of books by looking at the movies watched by the household occupants). However, it is important to build a generic platform that supports different types of privacy preserving data analytical capabilities. The UCN is expected to be made available to the public in 2017.

[1] J. Stankovic, "Research Directions for the Internet of Things," *Internet of Things Journal, IEEE,* vol. 1, no. 1, pp. 3-9, Feb 2014.

[2] D. McAuley, R. Mortier and J. Goulding, "The Dataware manifesto," in *Communication Systems and Networks (COMSNETS), 2011 Third International Conference on*, 2011.

**Author's Biography**

**Charith Perera** received his BSc (Hons) in Computer Science in 2009 from Staffordshire University, Stoke-on-Trent, United Kingdom and MBA in Business Administration in 2012 from University of Wales, Cardiff, United Kingdom and PhD in Computer Science at The Australian National University, Canberra, Australia. He is also worked at Information Engineering Laboratory, ICT Centre, CSIRO and involved in OpenIoT Project (FP7-ICT-2011.1.3) which is co-funded by the European Commission under seventh framework program. He has also contributed into several projects including EPSRC funded HAT project (EP/K039911/1)  His research interests include Internet of Things, Smart Cities, Mobile and Pervasive Computing, Context-awareness, Ubiquitous Computing. He is a member of both IEEE and ACM.

**Rajiv Ranjan** is a Senior Research and Julius Fellow at CSIRO, Canberra, where he is working on projects related to Cloud and big data computing. He has been conducting leading research in the area of Cloud and big data computing developing techniques for: (i) Quality of Service based management and processing of multimedia and big data analytics applications across multiple Cloud data centers (e.g., CSIRO Cloud, Amazon and GoGrid); and (ii) automated decision support for migrating applications to data centers. He has published about 110 papers that include 60+ journal papers. He serves on the editorial board of IEEE Transactions on Computers, IEEE Transactions on Cloud Computing, IEEE Cloud Computing, and Future Generation Computer System Journals. According to Google Scholar Citations his papers have received about 3450+ citations and he has an h-index of 24.

**Lizhe Wang** is a Professor at Institute of Remote Sensing & Digital Earth, Chinese Academy of Sciences (CAS) and a ChuTian Chair Professor at School of Computer Science, China University of Geosciences (CUG). Prof. Wang received his B.E. & M.E from Tsinghua Univ. and Doctor of Eng. from Univ. Karlsruhe, Germany. Prof. Wang is a Fellow of IET, Fellow of British Computer Society, and Senior Member of IEEE. Prof. Wang leads a group at CAS on Spatial Data Processing and the High Performance Computing Lab at CUG.

**Samee U. Khan** received a BS degree in 1999 from Ghulam Ishaq Khan Institute of Engineering Sciences and Technology, Topi, Pakistan, and a PhD in 2007 from the University of Texas, Arlington, TX, USA. Currently, he is Associate Professor of Electrical and Computer Engineering at the North Dakota State University, Fargo, ND, USA. Prof. Khan's research interests include optimization, robustness, and security of: cloud, grid, cluster and big data computing, social networks, wired and wireless networks, power systems, smart grids, and optical networks. His work has appeared in over 250 publications. He is on the editorial boards of leading journals, such as IEEE Transactions on Computers, IEEE Access, IEEE Cloud Computing Magazine, IEEE Communications Surveys and Tutorials, Scalable Computing, Cluster Computing, Security and Communication Networks, and International Journal of Communication Systems. He is a Fellow of the Institution of Engineering and Technology (IET, formerly IEE), and a Fellow of the British Computer Society (BCS). He is a Senior Member of the IEEE.

**Albert Y. Zomaya** is currently the Chair Professor of High Performance Computing & Networking in the School of Information Technologies, The University of Sydney. He is also the Director of the Centre for Distributed and High Performance Computing which was established in late 2009. Professor Zomaya is the author/co-author of seven books, more than 450 papers, and the editor of 14 books and 20 conference proceedings. He is the Editor in Chief of the IEEE Transactions on Computers and serves as an associate editor for 19 leading journals, such as, the ACM Computing Surveys and Journal of Parallel and Distributed Computing. Professor Zomaya is the recipient of the IEEE TCPP Outstanding Service Award and the IEEE TCSC Medal for Excellence in Scalable Computing, both in 2011. He is a Chartered Engineer, a Fellow of AAAS, IEEE, IET (UK).